\documentclass[conference, nofootinbib]{IEEEtran}
\usepackage{mdwtab}
\usepackage{cite}
\usepackage{notoccite}

\usepackage{comment}
\usepackage{color}
\usepackage[ampersand]{easylist}
\usepackage{times}
 \usepackage{graphicx}
\usepackage{booktabs}

\usepackage{listings}
\usepackage{bmpsize}
\usepackage{epstopdf}
\usepackage{array}
\usepackage{caption}
\usepackage{subcaption}
\usepackage{multicol}
\usepackage{fancyhdr}
\usepackage{lipsum}
\usepackage[english]{babel}
\usepackage{mathtools}
\usepackage[export]{adjustbox}
\usepackage{MnSymbol}
\usepackage{wasysym}
\usepackage{mdframed}
\newcolumntype{x}[1]{>{\centering\arraybackslash\hspace{0pt}}p{#1}}
\usepackage{array}
\usepackage[11pt]{moresize}

\newcolumntype{M}[1]{>{\centering\arraybackslash}m{#1}}
\newcolumntype{N}{@{}m{0pt}@{}}

\usepackage{balance}
\usepackage[export]{adjustbox}
\usepackage{epstopdf}
\usepackage{fixme}
\fxusetheme{color}
\fxsetup{
     status=draft,
     author=,
     layout=inline, 
     theme=color
}
\definecolor{fxnote}{rgb}{0.8000,0.0000,0.0000}

\newcommand{\B}[1]{\textbf{#1}}

\renewcommand{\footnoterule}{%
  \kern -3pt
  \hrule width \columnwidth height 0.1pt
  \kern 2pt
}

\begin{document}

\title{A Load-Balanced Parallel and Distributed Sorting Algorithm Implemented with PGX.D}

\author{$^{*}$Zahra Khatami$^{1,2}$, Sungpack Hong$^{1}$, Jinsoo Lee$^{1}$, Siegfried Depner$^{1}$, \\
Hassan Chafi$^{1}$, J. Ramanujam$^{2}$, and Hartmut Kaiser$^{2}$\\
$^{1}$Oracle Labs, USA\\
$^2$Center for Computation and Technology, Louisiana State University}
\maketitle

 \footnotetext[1] {This work was done during the author's internship at Oracle Labs.}

\begin{abstract}

Sorting has been one of the most challenging studied problems in different scientific researches. Although many techniques and algorithms have been proposed on the theory of having efficient parallel sorting implementation, however achieving desired performance on different types of the architectures with large number of processors is still a challenging issue. Maximizing parallelism level in applications can be achieved by minimizing overheads due to load imbalance and waiting time due to memory latencies. In this paper, we present a distributed sorting algorithm implemented in PGX.D, a fast distributed graph processing system, which outperforms the Spark's distributed sorting implementation by around 2x-3x by hiding communication latencies and minimizing unnecessary overheads. Furthermore, it shows that the proposed PGX.D sorting method handles dataset containing many duplicated
data entries efficiently and always results in keeping balanced workloads for different input data distribution types.

\end{abstract}


\section{Introduction}

Having efficient and scalable distributed sorting method has been one of the most studied problems in computer science, since it is of importance for many algorithms and also part of widely adopted benchmarks for parallel supercomputers \cite{s1,s2}. There have been many different parallel sorting techniques proposed, however, improving the scalability of a distributed sorting algorithm with conventional techniques is still one of the major challenges. Overall, the scalability of the parallel distributed sorting technique mostly depends on how well overheads, synchronizations and latencies are scheduled in both algorithm and the framework that the algorithm is implemented in. It is usually hard to achieve the optimum performance with existing frameworks since they often impose large overheads for scheduling, introduce unnecessary synchronization steps and are not able to handle large amount of memory consumption \cite{mpi1, mpi2, hpx1, hpx2}. In addition to these problems the system needs to partition the data entries between the machines equally, otherwise, it results in having workload imbalance, which in fact would prevent the system from scaling when dealing with large data. The communication overhead is another big challenge in these systems that needs to be taken in account when designing an efficient sorting technique.

In this research, we propose a new distributed sorting method, which overcomes these challenges by keeping balanced load and minimizes the overheads by fetching data efficiently in the partitioning and merging steps. The new handler is proposed that results in having a balanced merging while parallelizing merging steps, which improves the parallel performance. Moreover, the new investigator is proposed that results in keeping a balanced workloads among the distributed processors while dealing with dataset containing many duplicated data entries. This method is implemented in PGX.D, which is a scalable framework for various distributed implementations. PGX.D \cite{p1,p2} is a fast, parallel and distributed graph analytic framework that is able to process large graphs in distributed environments while keeping workloads well balanced among distributed machines. It improves the performance of the proposed sorting technique by exposing programming model that intrinsically reduces poor utilization of the resources by maintaining balanced workloads, minimizes latencies by managing parallel tasks efficiently and provides asynchronous task execution for sending/receiving data to/from the remote processors. The results presented in \cite{p1} show that PGX.D has low overhead and a bandwidth efficient communication framework, which easily supports remote data pulling patterns and is about 3x-90x faster than the other distributed graph systems such as GraphLab. Moreover, PGX.D decreases communication overheads by delaying unnecessary computations until the end of the current step, which allows the other processes to be continued without waiting for the completion of all the previous computations. Also it allows having asynchronous local and remote requests that avoids unnecessary synchronization barriers that helps in increasing scalability of the distributed sorting method \cite{p3}. 

In this paper, we show the performance improvements using distributed sorting method implemented in PGX.D to hide the communication latencies and to achieve better scalability. The performance of the proposed sorting framework shows around 2x-3x faster performance compared to the Spark's distributed sorting implementation.  Spark is a fast engine for processing large-scale data and it is widely used for processing large amount of data in parallel and distributed. Moreover, the results of the proposed technique confirm maintaining load balance regardless of the data distribution type.

The remainder of this paper is structured as follows: in section \ref{sec:related_work}, the related researches and the methods for distributed sorting are briefly explained; background information about PGX.D is provided in section \ref{sec:background_info}; the proposed PGX.D distributed sorting method and its implementation are explained in more details in section \ref{sec:model}, and the experimental results and the comparison between the proposed distributed sorting method with Spark's distributed sorting technique implemented for different data distributions are presented in section \ref{sec:results}. Conclusions can be found in section \ref{sec:conclusions}.

\section{Related Works}
\label{sec:related_work}

In this section, we briefly discuss about the implementation of some of the existing sorting techniques and their challenges. Different parallel algorithms have been studied for many years to achieve an efficient distributed sorting method implemented on distributed machines. Keeping balanced load is usually difficult while implementing most of these algorithms when facing dataset containing many duplicated data entries, which usually results in having some machines holding larger amount of data than the others. Most of these techniques consist of two main steps: \textit{(a)} partitioning, and \textit{(b)} merging. The big challenge in the partitioning step is to maintain the load balancing on different distributed machines in addition to distribute data among them in a sorting order \cite{s4,s7}. Also, the big challenge in the merging step is decreasing latencies and keeping the desired scalability during all merging steps till the end. The performance of the merging step is usually scalable when having a small number of machines. However, while utilizing larger number of machines, the synchronizations reduce its parallel scalability. Moreover, most of these existing techniques require a significant communication bandwidth, which increases overheads and degrades the parallel performance \cite{s5}.

Batcher's bitonic sorting \cite{bit3} is basically a parallel merge-sort and was considered as one of the most practical sorting algorithms for many years. In this algorithm, data on each local is sorted, then for each processors pair, the data sequence is merged into ascending order while on the other pair, the data sequence is merged into descending order. This algorithm is popular because of its simple communication pattern \cite{bit1,bit2}. However, it usually suffers from high communication overhead as its merging step highly depends on the data characteristics and it often needs to exchange the entire data assigned to each processor \cite{s6}.

Radix sort \cite{rad1} is also used for implementing parallel and distributed sorting algorithms in many different applications due to the simplicity of its implementation\cite{radix1}. This algorithm is not based on data comparison for  sorting purposes. It is implemented by considering bit representation of the data. The algorithm starts by examining the least significant bit in each data entry. Then, in each time step, data entries are being sorted by considering their r-bit values. One of the big challenges in implementing this sorting technique is having unequal number of input keys\cite{merge,s2}. It usually suffers in irregularity in communication and computation. Since it also highly depends on the data characteristics, the computation on each processor and the communication between different processors are not known in the very first steps, which can negatively affect its parallel performance. 

In order to increase the parallel performance of the sorting algorithm, its structure and design should be independent of the data characteristics to decrease communication overheads. Therefore, sample sort is often chosen for implementing distributed sorting. It is a fast sorting algorithm that keeps the load balancing better than the quick sort and doesn't have the mentioned communication problem faced in Batcher's bitonic sort and radix sort\cite{sample,sample2}. The runtime in this algorithm is almost independent of the input data distribution types. It works by choosing some samples regularly from the locally sorted data of each processor and selecting final splitters from these samples to split data into different number of processors for the sorting purpose. However, in addition to having scalability challenge in its merging step, its performance highly depends on choosing sufficient amount of samples from each processor and picking the efficient final splitters from them. If this step is not efficiently designed, the algorithm may result in having load imbalance when facing dataset containing many duplicated data entries. 

Nowadays, \textit{Spark} is widely used for processing large data as it provides the scalable framework for analyzing large input data on distributed machines. Spark provides a resilient distributed dataset (RDD), which is a collection of tasks that are partitioned across distributed machines and they can be processed in parallel. The distributed sorting algorithm implemented in Spark uses three main stages: sample, map and reduce\cite{spark1, spark2}. In the sample stage, samples from the data is chosen for having balanced data partitioned among the distributed machines. In the map stage, data is partitioned on distributed machines and is locally sorted and being merged in the reduce stage. TimSort \cite{Tim1} is chosen as a sorting technique in Spark and the experimental results show that it performs better when the data is partially sorted. This algorithm starts by finding subsequences of the elements in  descending or ascending order and performs balanced merges on them in each merging step. For this purpose, it proceeds on the chosen minimum run sizes that are bulked up by using insertion sort and partially merge them in place, which helps its implementation to fetch data efficiently. 

In this paper, the sample sorting technique is chosen as the base method for implementing a distributed sorting technique in PGX.D, which solves the mentioned challenges by implementing improvement in choosing splitters and managing communications between different processors efficiently. In addition, the performance optimizations considered in TimSort are also applied in the proposed sorting technique. A sorting library is generated in PGX.D that performs efficiently by:

\begin{enumerate}
\item[] $\checkmark$ maintaining load balance on different number of processors, even when having a dataset containing many duplicated data entries,

\item[] $\checkmark$ reducing overheads in the merging step by keeping balanced merging and fetching data efficiently,

\item[] $\checkmark$ reading/writing data from/to the remote processors asynchronously, which avoids the unnecessary synchronizations in each step, and

\item[] $\checkmark$ keeping bandwidth-efficient communication implemented in PGX.D.
\end{enumerate}

Our sorting technique is modeled by considering problem size, number of processors and type of the data distribution for improving load balancing. Section \ref{sec:background_info} discusses how PGX.D framework helps distributed sorting to reduce the synchronizations and the latencies during the sorting implementation. The programming model of the distributed sorting method is explained in more details in section \ref{sec:model}.

\section{Background Information}
\label{sec:background_info}

In this section the overview of PGX.D is presented to show its capability in providing scalable framework for implementing efficient distributed sorting. PGX.D is a fast, parallel and distributed graph analytic framework and it is able to process large graphs in distributed environments while keeping workloads well balanced among multiple distributed machines \cite{p1,p2}. It uses a fast context switching mechanism that results in having low overhead in task scheduling as well as a bandwidth efficient communication process. It guarantees low communication overhead by applying ghost nodes selection that results in decreasing number of the crossing edges as well as decreasing communication between different processors. Also, a new edge chunking strategy is implemented that improves task scheduling and results in having balanced workload between the processors in each machine. Moreover, it relaxes the unnecessary synchronizations while applying local/remote write requests. Figure \ref{f1} shows the architecture of PGX.D. The layers of the system are explained as follows:

\begin{figure}
\begin{center}
\centering
\includegraphics[width=0.8\columnwidth]{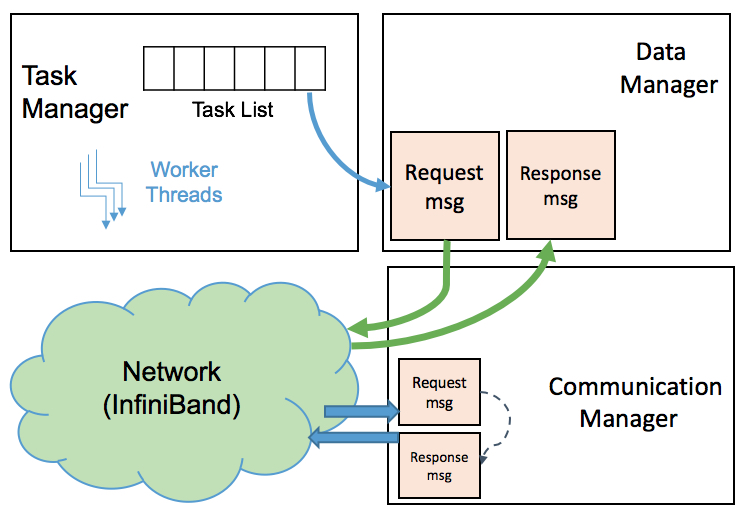}
\caption {\small{Three important mangers in PGX.D}}
\label{f1}
\end{center}
\end{figure}

$\bullet$ Task Manager: A list of tasks is created within a task manager at the beginning of each parallel step. The task manager initializes a set of worker threads and each of these threads grab a task from the list and executes it. If the task is requested from the remote machine, the execution of the process is continued until the data becomes available, which relaxes the synchronization barriers. Also, it determines the schedules for sending messages to the remote machines by considering size of the messages stored in the request buffer and sending them whenever that buffer reaches to its maximum size or the worker thread has completed all its scheduled tasks. 

$\bullet$ Data Manger: Graph data across different machines is maintained within the data manager and they are stored in the Compressed Sparse Row (CSR) data structure on each machine. Whenever a message has to be sent to the remote machine, the data manager buffers it up in the request buffers for the task manager while keeping corresponding data structure. Also, the partitioning step, choosing ghost nodes and edge chunking on the graph data is applied within this manager during the graph loading. These implementations reduce communication overheads significantly and result in improving load balancing \cite{p1}. The location of each node is identified with this manager as well, which allows to handle a data transformation and to determine the destined machine for the messages within the request buffers.

$\bullet$ Communication Manager: The response messages to the remote machines are managed with the communication manager. The main goal of the 
communication manager is to provide communication primitives which allow for low overhead communication. It provides a fast and a low overhead mechanism while processing incoming request messages and applies the atomic instructions while processing writing requests. Also, it is able to handle large amount of small sized communication between different machines efficiently.

PGX.D takes care of the synchronizations between each step and provides both write and read requests within the same parallel region. As a result, PGX.D framework is more relaxed compared to the bulk-synchronization model used in the MapReduce models. The previous analysis and experiments show a good performance on the certain types of the graph algorithms. For more details about PGX.D, we recommend the reader to refer to \cite{p1}. In this research, we choose PGX.D to implement the distributed sorting techniques, which helps the algorithm to perform efficiently and scales desirably. Moreover, by adding this distributed sorting method in PGX.D, user can also easily sort data of their multiple graphs with different types and implement more analysis on them, such as retrieving top values from their graph data or implementing binary search on the sorted data. The programming model of the PGX.D distributed sorting technique is explained in more details in the next section.

\section{Programming Model}
\label{sec:model}

The PGX.D distributed sorting technique is designed based on sample sort algorithm, which is one of the most popular sorting algorithms and is usually faster than than other techniques by better maintaining load balance and minimizing communication overheads \cite{sample,sample2}. As discussed in section \ref{sec:related_work}, this algorithm works by choosing samples from the locally sorted data of each processor and selecting final splitters from these samples to split data into different number of processors for the sorting purpose. However, one of the big challenges that can easily impact the performance of this sorting algorithm is having non-efficient splitters. In other words, if the splitters do not spilt data into the equal pieces, applying this algorithm may result in having larger amount of data on some processors compared to the others, which can degrade the performance significantly. Therefore this sampling step can be considered as the critical step for this algorithm that has received a wide attention for the development to obtain the expected workload on distributed processors. The performance of this step depends on the number of samples received on Master from the other processors, which arises two different scenarios that should be considered:

$\B{A)}$ If a large amount of samples is sent to Master from each processor, Master will get a good idea about a data distribution on each processor, which helps it to choose the efficient splitters from the received samples that results in having better load balance. However, sending large amounts of data from each processor to Master, increases communication overheads as well as increasing the time needed for merging received data on Master.

$\B{B)}$ If a small amount of samples is chosen from each processor, the communication overheads will be decreased. However, Master may be unable to choose splitters which cause an equal distribution which results in having load imbalance issues. 

PGX.D distributed sorting method manages this issue by designing scalable framework that efficiently uses the bandwidth efficient as well as providing large enough number of samples for keeping balanced workloads. In addition, handling balanced merging and providing asynchronous task executions in sending/receiving data to/from the other processors are another features that are provided with the proposed sorting technique in PGX.D to achieve a desired parallelism level. These features are considered while implementing proposed sorting technique that mainly consists of six steps, which will be briefly outlined here and explained in more detail later.


\begin{enumerate}

\item Each processor sorts its data locally by using parallel quick sort.

\item The regular samples are chosen from the sorted data and they are sent to Master.

\item Master selects the final splitters from the received samples and it broadcasts them to all processors.

\item Each processor determines the range of the data to be sent to the remote processor by implementing binary search of these splitters on their local sorted data.

\item After setting up the senders and the receivers, each processor starts sending and receiving data simultaneously.

\item After receiving data from different remote processors, all data is merged together while keeping information regards to their previous processors and locations.

\end{enumerate}

\subsection{Handling Balanced Merging}

\begin{figure}
\begin{center}
\centering
\includegraphics[width=0.8\columnwidth]{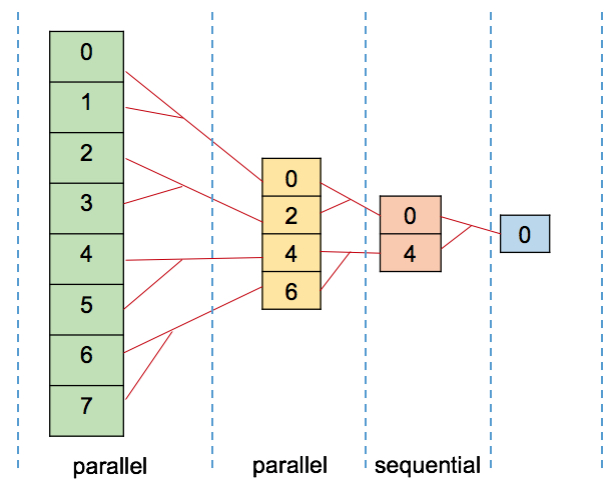}
\caption {\small{Keeping balanced merging in PGX.D distributed sorting method.}}
\label{f2}
\end{center}
\end{figure}

In step $\B{(1)}$, data is sorted locally on each processor by implementing parallel quick sort in such a way that data is divided equally among a number of the worker threads on each processor. Then, each worker thread sorts its data locally. Sorted data from each thread is merged together by keeping balanced merging. For this purpose, the new handler is proposed in this section that manages the merging implementation to keep the balanced merging in each step, which avoids the cache misses. For example in figure \ref{f2} with $8$ threads, in the first step, the data in thread $1$ is merged with the data in thread $0$, the data in thread $3$ is merged with the data in thread $2$, the data in thread $5$ is merged with the data in thread $4$, and the data in thread $7$ is merged with the data in thread $6$. So, at the end of the first step, data on the threads with odd ID are merged with data on the threads with even ID and they have almost the equal sizes. In the next step, the data in thread $2$ is merged with the data in thread $0$ and the data in thread $6$ is merged with the data in thread $4$. In the last step, the data in thread $4$ is merged with the data in thread $0$. Also, this handler implements each of these merging steps in parallel which improves the parallel performance of this sorting technique.

\subsection{Providing Large Enough Number of samples}

PGX.D distributed sorting method is designed in such a away that it keeps the communication overheads as less as possible while letting the largest possible amount of samples to be sent from each processor destined to the Master in step $\B{(2)}$ to achieve a better load balance. The amount of data is chosen by considering number of processors and the size of a read buffer defined in the PGX.D data manager. The size of this buffer is assigned $256$ Kbyte in PGX.D based on measuring different performances and choosing the best one \cite{p1}. So if there are $p$ number of processors in the system, each processor has to send only $256 / p$  Kbyte data to Master, which results in sending only one buffer data entries from each processor to Master and gathering all received data entries in one receiving buffer on Master. The experimental results of this algorithm on different types of data confirm that the overheads are significantly reduced and the amount of data provided to the Master is large enough to choose the efficient splitters.

\begin{figure}
\centering

\begin{subfigure}{0.5\textwidth}
  \centering
  \includegraphics[width=1\columnwidth]{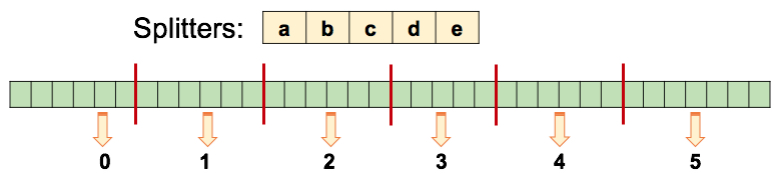}
  \caption{\small{Non-duplicated data\\}}
  \vspace*{5mm}
  \label{f5_a}
\end{subfigure}

\hfill

\begin{subfigure}{0.5\textwidth}
  \centering
  \includegraphics[width=1\columnwidth]{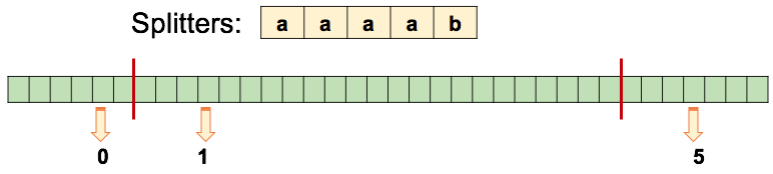}
  \caption{\small{Duplicated data}}
    \vspace*{5mm}
  \label{f5_b}
\end{subfigure}

\begin{subfigure}{0.5\textwidth}
  \centering
  \includegraphics[width=1\columnwidth]{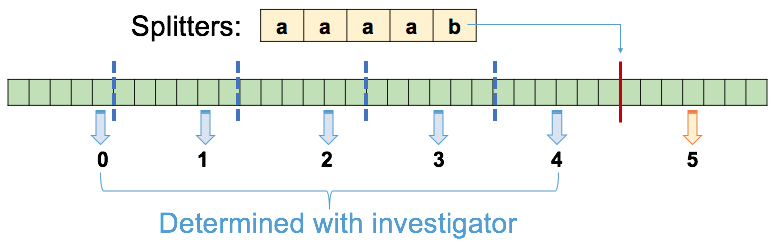}
  \caption{\small{Dividing range between duplicated splitters}}
  \label{f5_c}
\end{subfigure}

\caption{\small{Determining range of data to be sent to the remote processor by implementing binary search for the received splitters on locally sorted data.}}
\label{fig:test}
\end{figure}

After receiving $p - 1$ splitters from step $\B{(3)}$, in step $\B{(4)}$, the ranges of data for sending to the other processors are determined by implementing binary search on data for each splitter. Then data from the splitter$[j - 1]$ to the splitter$[j]$ is sent to the processor $j - 1$, as it is illustrated in figure \ref{f5_a}. However, if there exists dataset containing many duplicated data entries on the processors, this method results in having  load imbalance by sending large data range to one processor and not sending any data to the others as shown in figure \ref{f5_b}. To solve this problem, we propose the investigator to be implemented during searching for each splitter on local data entries, which makes a binary search to be executed for only non-duplicated splitters, then it divides the determined range between the duplicated splitters equally. For example in figure \ref{f5_c}, the range is divided equally between $4$ duplicated splitters with the same \textit{a} value. 

\subsection{Asynchronous Task Execution} 

After determining ranges for each destination in step $\B{(4)}$, these information are broadcasted to all processors. So each processor knows how much data it will receive from the other processors. This strategy enables the sorting technique to receive data from multiple processors at the same time by applying offsets for each received data entry from the neighbor lists, which allows writing in the local data list simultaneously. Also each processor is able to send data while receiving data, which avoids the unnecessary synchronizations between these steps as well. In the last step, the new arriving data is merged together in parallel by applying same method described in figure \ref{f2}. 

At the end, data is sorted in such a way that smaller data entries is stored and gathered in the processor with the smaller ID and the bigger data is stored and gathered in the processor with the bigger ID. This sorting library also provides an API for the users to implement a binary search on data as well as finding information regards to the previous processors and the previous indexes of the new received data entry of each processor. Also it is able to sort multiple different data simultaneously and keeping the balanced load while having a dataset containing many duplicated data entries. The experimental results of PGX.D's distributed sorting technique is studied in more details in section \ref{sec:results}.

\section{Experimental Results}
\label{sec:results}

\begin{table}
\centering
\begin{tabular}{lll}
       \toprule
       Category & Item & Detail \\
       \hline
      SW & OS & Linux 2.6.32\\
       & compiler & gcc 4.8.2\\
      \hline
       CPU & Type & Intel Xeon E5-2660 \\
        & Frequency & 2.40 GHz\\
        & Parallelism & 2 socket with 8 cores \\
       \hline
       DRAM & Type &  DDR3-1600 \\
       & Size & 256 GB\\
       \hline
       Size & $\#$ Machines  & 32\\
       \hline
       Network & Card & Mellanox Connect-IB\\
       & Switch & Mellanox SX6512\\
       & Raw BW & 56Gb/s (per port)\\
       \bottomrule
       
    \end{tabular}
    \caption{\small{Experimental Environment}}
    \label{t1}
\end{table}

\begin{figure}
\centering

\begin{subfigure}{0.21\textwidth}
  \includegraphics[width=\textwidth]{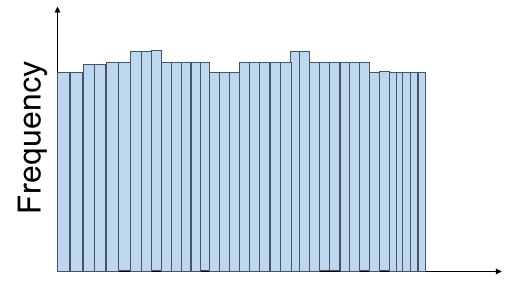}
 \caption{\small{Uniform distribution}}
  \label{f6_a}
\end{subfigure}
\begin{subfigure}{0.21\textwidth}
  \includegraphics[width=\textwidth]{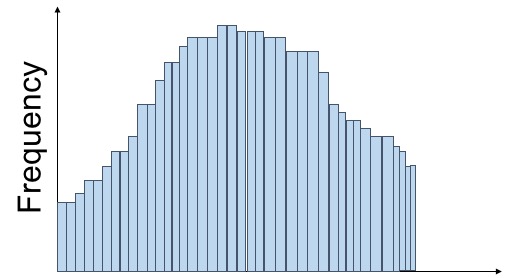}
  \caption{\small{Normal distribution}}
  \label{f6_b}
\end{subfigure}

\begin{subfigure}{0.21\textwidth}
  \includegraphics[width=\textwidth]{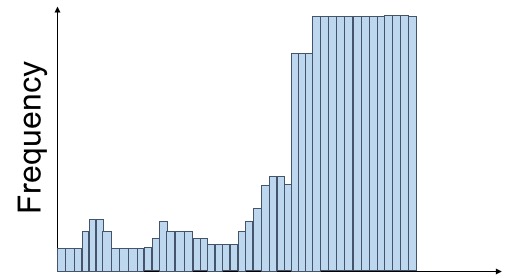}
  \caption{\small{Right-skewed distribution}}
  \label{f6_c}
\end{subfigure}
\begin{subfigure}{0.21\textwidth}
  \includegraphics[width=\columnwidth]{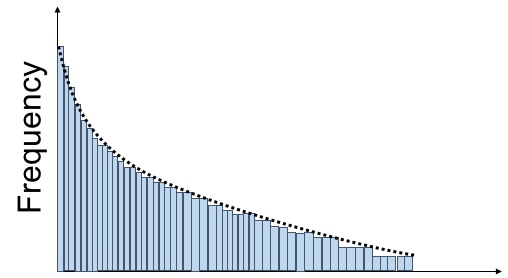}
  \caption{\small{Exponential distribution}}
  \label{f6_d}
\end{subfigure}

\caption{\small{Different types of data distribution}}
\label{fig:test}
\end{figure}

The proposed PGX.D distributed sorting mechanism has been implemented for different types of data on our cluster, which is summarized in Table \ref{t1}. Its performance is compared with the Spark's distributed sorting implementation with version 1.6.1 that is publicly available. For our evaluation, first we study the proposed sorting technique performance on four different input datasets shown in figure \ref{fig:test} with the uniform, normal, right-skewed, and exponential distribution type. The implementation of our proposed sorting technique on  data with the right-skewed and the exponential distribution type are specially intended to confirm its ability to maintain load balancing in a case of having a dataset containing many duplicated data entries. All of these input datasets have one billion entries and the sorting is performed on $8$ up to $52$ processors with using $32$ threads per each for in-node parallelization purposes. Distributed sorting method in Spark is executed by implementing \textit{sortByKey()} in the MapReduce operation.

\begin{figure}
\begin{center}
\centering
\includegraphics[width=\columnwidth]{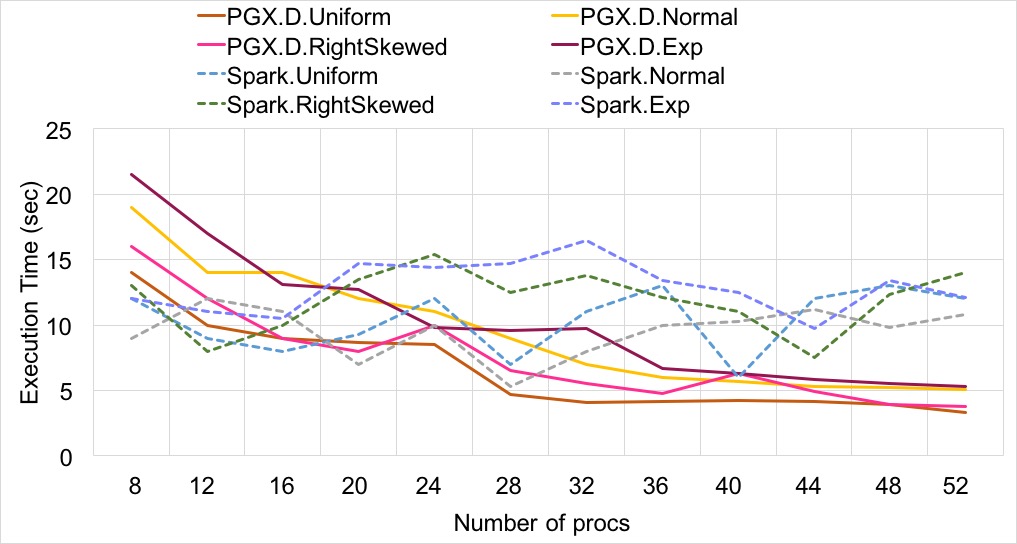}
\caption {\small{PGX.D distributed sorting method total execution times for data from figure \ref{fig:test}.}}
\label{f5}
\end{center}
\end{figure}

\begin{figure}
\begin{center}
\centering
\includegraphics[width=\columnwidth]{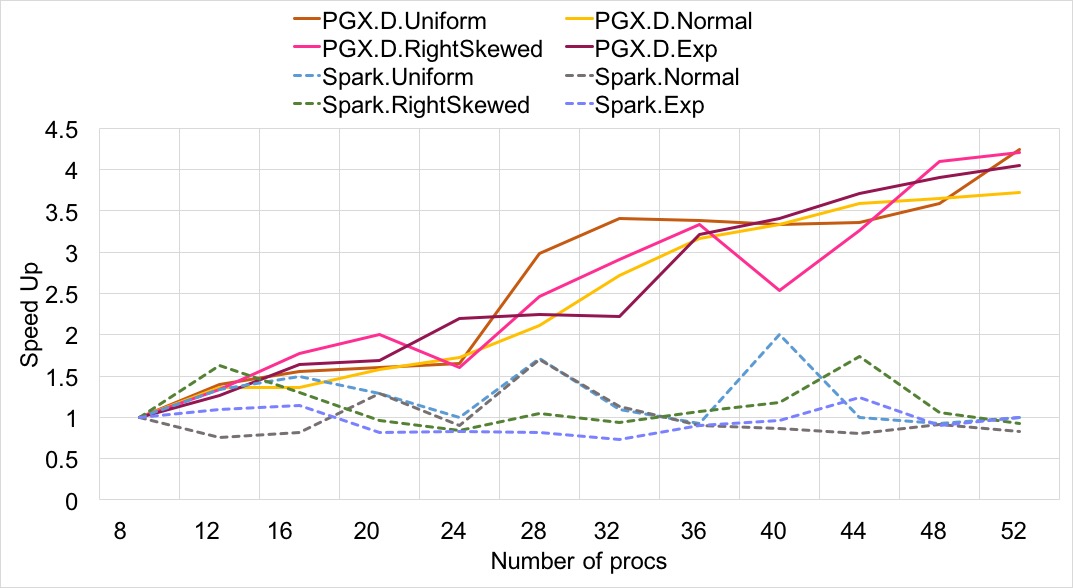}
\caption {\small{Comparison results of strong scaling of the distributed sorting method implemented in PGX.D and Spark for data from figure \ref{fig:test}.}}
\label{f6}
\end{center}
\end{figure}

Figure \ref{f5} shows the execution time of the distributed sorting methods on data from figure \ref{fig:test}. It illustrates that PGX.D sorts data efficiently regardless of the input data distribution type. Since the communication latencies between different processors increase with an increasing number of processors, it is usually hard to achieve a good scaling behavior. However, figure \ref{f6} shows a better speedup of PGX.D distributed sorting technique compared to the sorting technique in Spark.

\begin{figure}
\begin{center}
\centering
\includegraphics[width=\columnwidth]{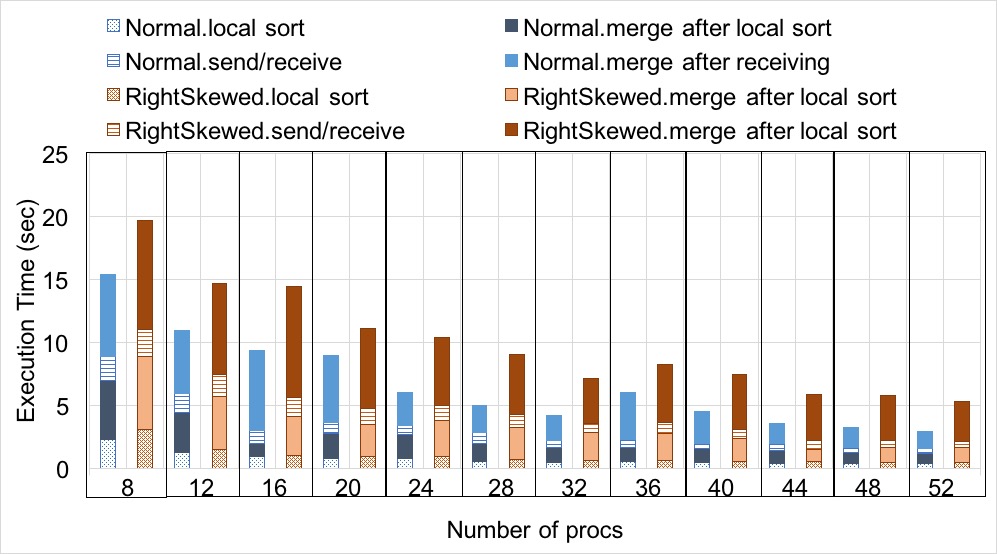}
\caption {\small{PGX.D distributed sorting method execution times of each different steps for the normal and right skewed distribution types from figure \ref{fig:test}.}}
\label{f5b}
\end{center}
\end{figure}

Figure \ref{f5b} shows the execution time of each steps for the experiments on the normal and right skewed distribution types studied in figure \ref{f5}. It can be seen that sending/receiving data costs less time than the other steps, which validates the efficient-bandwidth communication and the asynchronous execution provided in PGX.D that remove the unnecessary barriers and allow to have a low communication overhead to achieve a better parallelization performance.

\begin{table*}[t]
\begin{minipage}{\textwidth}
\centering
    \begin{tabular}{|c|c|c|c|c|c|c|c|c|c|c|}
       \hline
       Distribution type & proc0 & proc1 & proc2 & proc3 & proc4 & proc5 & proc6 & proc7 & proc8 & proc9\\
       \hline
      Uniform &  9.984\% & 9.989\% & 9.998\% & 10.003\% & 10.022\% & 10.000\% & 9.993\% & 9.994\% & 10.018\% & 9.993\% \\
      \hline
       Normal  & 10.011\%  & 10.003\% & 10.000\% & 10.000\% & 9.981\% & 9.999\% & 9.992\%  & 10.008\% & 9.999\% & 10.002\% \\
       \hline
       Right-skewed  & 10.020\% & 9.990\% & 9.998\% & 9.998\% & 9.998\% & 9.998\% & 9.998\% &  9.998\% & 9.998\% &  9.998\%\\
       \hline
       Exponential  &  9.998\% & 9.998\% & 9.998\% & 9.998\% &  9.997\% & 9.997\% & 9.997\%  &  9.996\% & 9.996\% & 10.020\% \\
    \hline
    \end{tabular}
    \caption{\small{The ratio of data sizes on each processor after implementing balanced PGX.D distributed sorting method using $10$ processors on $1$ billion data from figure \ref{fig:test}.}}
    \label{t2}
    \end{minipage}
\end{table*}

Table \ref{t2} shows the size of data on each processor after PGX.D distributed sorting implementation having $10$ processors. It illustrates data is distributed equally on the processors, in the case of having a dataset containing many duplicated data entries in both right-skewed and exponential distribution types. Also, these information generally shows how the proposed investigator works during a binary search implementation on the duplicated splitters. For example, the results according to the sizes of data in the right-skewed distribution show having the exact equal sized $9.998\%$ for each data on the processors $2 - 9$. 

\begin{figure}
\begin{center}
\centering
\includegraphics[width=\columnwidth]{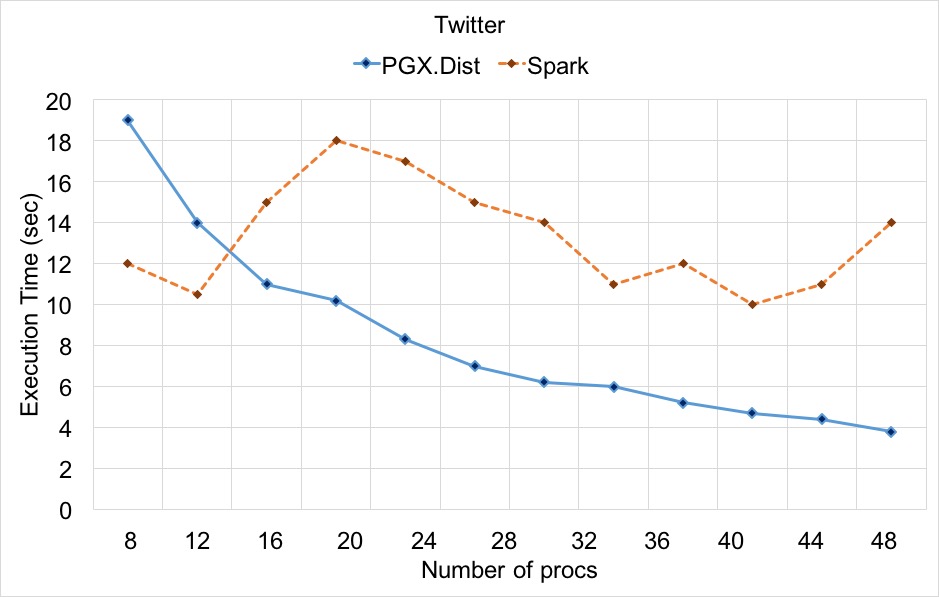}
\caption {\small{Execution time of distributed sorting method implemented in PGX.D compared to Spark's distributed sorting implementation on Twitter graph dataset.}}
\label{f8}
\end{center}
\end{figure}

We also evaluate the performance of our sorting technique on the Twitter graph dataset, which has $41.6$ million vertices and $1.5$ million edges in a $25$GB dataset. Figure \ref{f8} shows the execution time compared to Spark's distributed sorting implementation, which illustrates that it is faster than Spark by around $2.6$x on $52$ processors. The ranges of data on each processor after sorting with $8$, $12$ and $16$ processors are included in Table \ref{t3}, which confirms the accuracy of the proposed technique that data with the smaller value are located on the processor with the smaller ID and data with the bigger value are located on the processor with the bigger ID. 

\begin{figure}
\begin{center}
\centering
\includegraphics[width=\columnwidth]{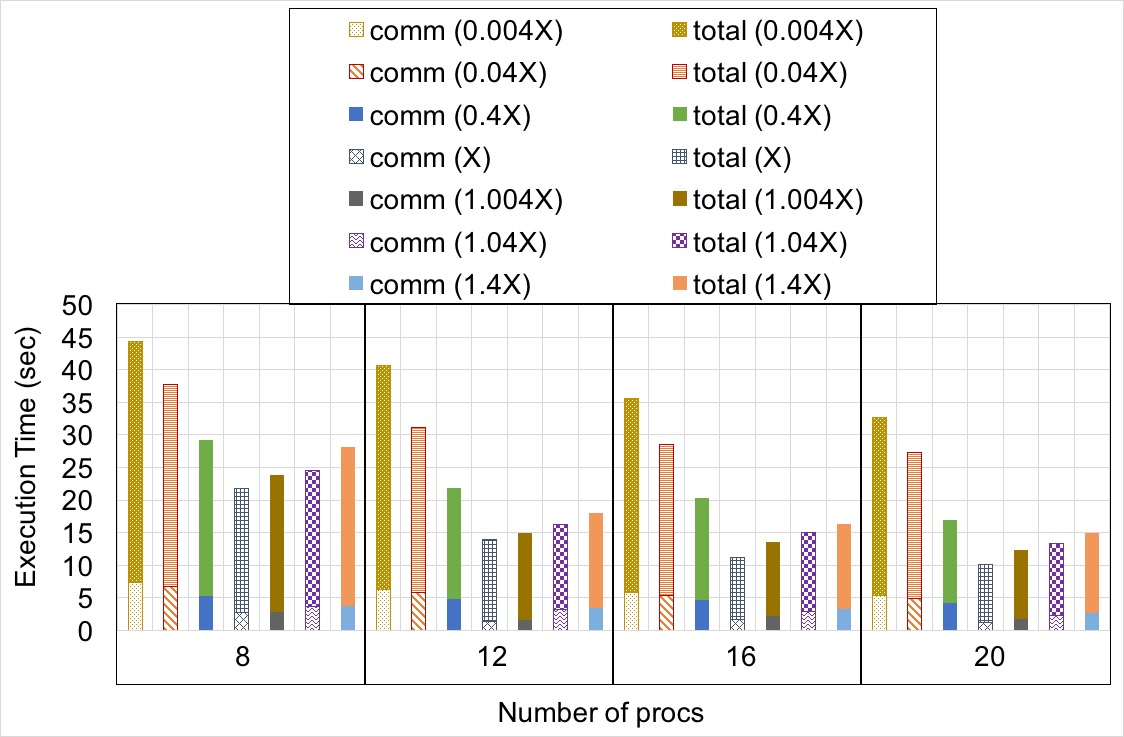}
\caption {\small{Communication overhead and total execution time comparison of distributed sorting method implemented for Twitter graph dataset using different sample sizes, where \textit{X = 256KB/number of processors}.}}
\label{f10}
\end{center}
\end{figure}

In figure \ref{f10}, we study about the impact of sample sizes on the communication overheads and the total execution time. Seven different sample sizes are used: $0.004X$, $0.04X$, $0.4X$, $X$, $1.004X$, $1.04X$, and $1.4X$, where \textit{X = 256KB/number of processors} (\textit{256KB} is equal to the size of the read buffer in PGX.D). It shows that the small number of samples not only results in having load imbalance, but it also increases communication overheads, which is due to having non-efficient splitters that result in sending more data to some number of processors. Also, the total execution time for the cases of having very small amount of samples and large amount of samples are both greater than the execution time of having $X$ samples. 

\begin{figure}
\begin{center}
\centering
\includegraphics[width=\columnwidth]{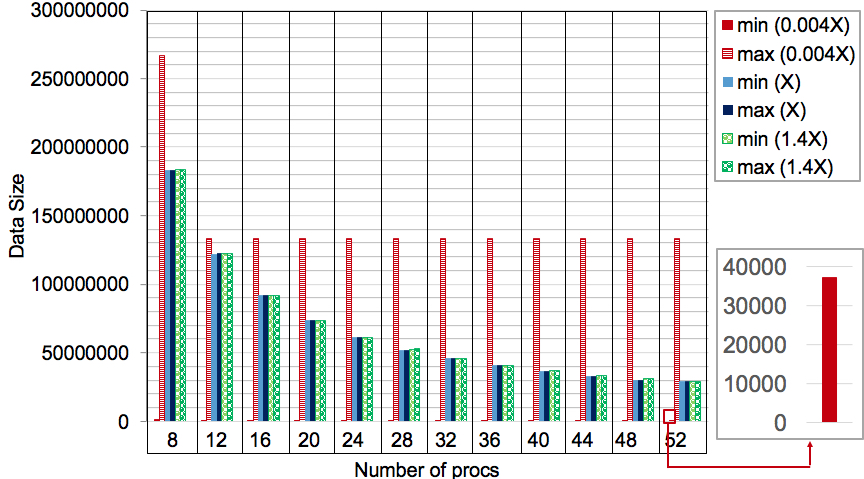}
\caption {\small{Work load comparison on each processor using different sample sizes on Twitter graph dataset, where \textit{X = 256KB/number of processors} .}}
\label{f9}
\end{center}
\end{figure}

In figure \ref{f9}, we study about the impact of sample sizes on the load balance from figure \ref{f10} for three different samples sizes as $0.004X$, $X$, and $1.4X$. The minimum and the maximum size of data is illustrated in using different number of processors. It shows that $0.004X$ number of samples is not large enough to keep balanced workloads between the processors, as it results having the load difference by an average $133192278$ while using $52$ processors. However, both $X$ and $1.4X$ result in having balanced loads in all experiments. By considering results from the figures \ref{f10} and \ref{f9}, $X$ is chosen as the sample size in the distributed sorting method implemented in PGX.D, which results in having load balance while keeping low communication overhead. 

\begin{figure}
\begin{center}
\centering
\includegraphics[width=\columnwidth]{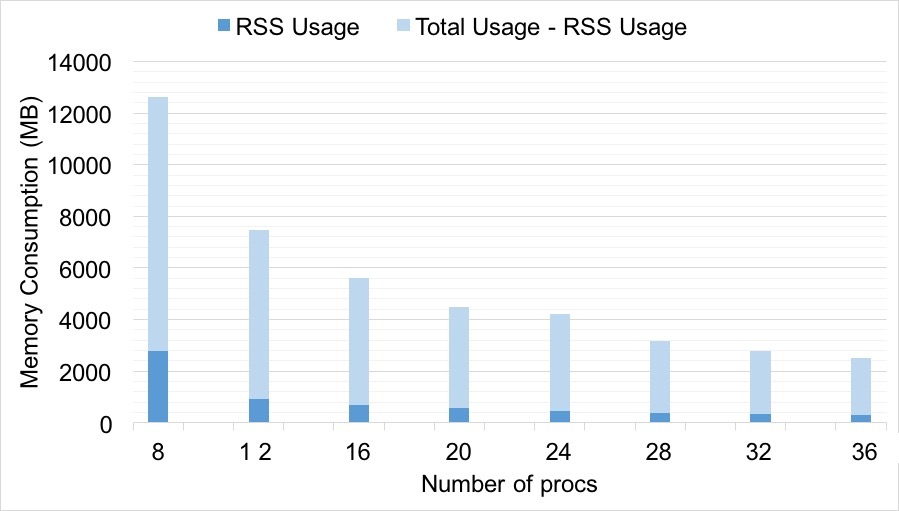}
\caption {\small{Average memory consumptions of PGX.D distributed sorting method on Twitter graph dataset using different number of processors.}}
\label{f7}
\end{center}
\end{figure}

Since the main purpose of this sorting technique is to distribute and to sort large amount of data efficiently, it is very important to avoid using extra memory during the implementation. The average memory consumptions using different number of processors for Twitter graph dataset are shown in figure \ref{f7}. Resident Set Size (RSS) is the RAM memory that is allocated for the process and it is shown in dark blue. Light blue illustrates the total temporary memory usage during the process except RSS usage, which is allocated during the process and becomes free at the end. It is shown that the maximum memory needed for sorting on $20$ processors is less than $300$MB, which is used for keeping previous information of each data's previous processor and location. 

\begin{table}
\centering
    \begin{tabular}{|c|c|c|c|}
        \hline
      $\#$ procs &8&12&16\\
       \hline
      proc0 & 0 - 12.58 & 0 - 8.39 & 0 - 6.29\\
      \hline
      proc1 & 12.58  -  25.07 & 8.39 - 16.68 & 6.29 - 12.58\\
      \hline
      proc2 & 25.07 - 37.55& 16.68 - 25.07 & 12.58 - 18.78\\
      \hline
      proc3 & 37.55 - 50.04& 25.07 - 33.36 & 18.78 - 25.06\\
      \hline
      proc4 & 50.04 - 62.53& 33.36 - 41.66 & 25.06 - 31.26\\
      \hline
      proc5 & 62.53 - 75.02& 41.66 - 50.04 & 31.26 - 37.55\\
      \hline
      proc6 & 75.02 - 87.50& 50.04 - 58.33 & 37.55 - 43.75\\
      \hline
      proc7 & 87.50 - 95& 58.33 - 66.63 & 43.75 - 50.04\\
      \hline
      proc8 &  & 66.63 - 75.02 & 50.04 - 56.24\\
      \hline
      proc9 & & 75.02 - 83.31 & 56.24 - 62.52\\
      \hline
      proc10 & & 83.31 - 91.61 & 62.52 - 68.73\\
      \hline
     proc11 & & 91.61 - 95 & 68.73 - 75.01\\
      \hline
      proc12 & & & 75.01 - 81.21\\
      \hline
      proc13 &&& 81.21 - 87.50\\
      \hline
      proc14 && & 87.50 - 93.70\\
      \hline
      proc15 & & & 93.70 - 95\\
     \hline
    \end{tabular}
    \caption{\small{Range of data on each processors after implementing balanced PGX.D distributed sorting method on Twitter graph dataset.}}
    \label{t3}
\end{table}

\section{Conclusions}
\label{sec:conclusions}

Implementing sample sort is straightforward and simple and can be faster than the other techniques if it is designed efficiently. In another words, despite its simplicity, it causes a load imbalance and increases  communication overheads if it is not able to handle having a dataset containing many duplicated data entries and if it is performed within a framework without the capability of exploiting bandwidth efficiently.  In this paper, we have developed a distributed sorting technique with PGX.D that eliminates the load balancing problem when sorting dataset containing many duplicated data entries on distributed machines. This technique is implemented by considering number of processors and the maximum buffer size provided with PGX.D, which helps gathering enough information from different machines without increasing communication overheads. Also, the new handler is proposed that results in having a balanced merging while parallelizing merging steps, which improves the parallel performance. Moreover, this technique is able to keep the workload balanced among the distributed processors by applying proposed investigator during a binary search. One of the other advantages of this sorting technique is the high-level API exposed to the user, which is a generic and works with any data type and is able to sort different data simultaneously.

To evaluate the performance of the proposed distributed sorting technique, we have measured it with different input datasets having various distribution types: uniform, normal, right-skewed and exponential distributions. We were able to sort one billion data on up to $52$ machines by maintaining load balance on all processors. In addition, a balanced PGX.D distributed sorting method has been compared with Spark's distributed sorting implementation on the same input data and the experimental results illustrated that our technique is by around $2x-3x$ faster than Spark's distributed sorting method. Also, it was shown that it uses as less memory as possible to avoid increasing  memory usage while facing big data. 


\bibliography{References}

\begin{thebibliography}{10}

\bibitem{s1}
Michael~C Loui.
\newblock The complexity of sorting on distributed systems.
\newblock {\em Information and Control}, 60(1):70--85, 1984.

\bibitem{s2}
Zhang Qian, Ge~Yufei, Liang Hong, and Shi Jin.
\newblock A load balancing task scheduling algorithm based on feedback
  mechanism for cloud computing.
\newblock {\em International Journal of Grid and Distributed Computing},
  9(4):41--52, 2016.

\bibitem{mpi1}
Rolf Rabenseifner, Georg Hager, Gabriele Jost, and Rainer Keller.
\newblock {Hybrid MPI and OpenMP parallel programming}.
\newblock In {\em PVM/MPI}, page~11, 2006.

\bibitem{mpi2}
Ashay Rane and Dan Stanzione.
\newblock {Experiences in tuning performance of hybrid MPI/OpenMP applications
  on quad-core systems}.
\newblock In {\em Proc. of 10th LCI Int'l Conference on High-Performance
  Clustered Computing}, 2009.

\bibitem{hpx1}
Zahra Khatami, Hartmut Kaiser, and J~Ramanujam.
\newblock Using hpx and op2 for improving parallel scaling performance of
  unstructured grid applications.
\newblock In {\em Parallel Processing Workshops (ICPPW), 2016 45th
  International Conference on}, pages 190--199. IEEE, 2016.

\bibitem{hpx2}
Zahra Khatami, Hartmut Kaiser, Patricia Grubel, Adrian Serio, and J~Ramanujam.
\newblock A massively parallel distributed n-body application implemented with
  hpx.
\newblock In {\em Proceedings of the 7th Workshop on Latest Advances in
  Scalable Algorithms for Large-Scale Systems}, pages 57--64. IEEE Press, 2016.

\bibitem{p1}
Sungpack Hong, Siegfried Depner, Thomas Manhardt, Jan Van Der~Lugt, Merijn
  Verstraaten, and Hassan Chafi.
\newblock Pgx. d: a fast distributed graph processing engine.
\newblock In {\em Proceedings of the International Conference for High
  Performance Computing, Networking, Storage and Analysis}, page~58. ACM, 2015.

\bibitem{p2}
Raghavan Raman, Oskar van Rest, Sungpack Hong, Zhe Wu, Hassan Chafi, and Jay
  Banerjee.
\newblock Pgx. iso: parallel and efficient in-memory engine for subgraph
  isomorphism.
\newblock In {\em Proceedings of Workshop on GRAph Data management Experiences
  and Systems}, pages 1--6. ACM, 2014.

\bibitem{p3}
Oskar van Rest, Sungpack Hong, Jinha Kim, Xuming Meng, and Hassan Chafi.
\newblock Pgql: a property graph query language.
\newblock In {\em Proceedings of the Fourth International Workshop on Graph
  Data Management Experiences and Systems}, page~7. ACM, 2016.

\bibitem{s4}
Minsoo Jeon and Dongseung Kim.
\newblock Parallelizing merge sort onto distributed memory parallel computers.
\newblock In {\em International Symposium on High Performance Computing}, pages
  25--34. Springer, 2002.

\bibitem{s7}
Shmuel Zaks.
\newblock Optimal distributed algorithms for sorting and ranking.
\newblock {\em IEEE Transactions on computers}, 100(4):376--379, 1985.

\bibitem{s5}
KVS Ramarao.
\newblock Distributed sorting on local area networks.
\newblock {\em IEEE Transactions on Computers}, 37(2):239--243, 1988.

\bibitem{bit3}
Yong~Cheol Kim, Minsoo Jeon, Dongseung Kim, and Andrew Sohn.
\newblock Communication-efficient bitonic sort on a distributed memory parallel
  computer.
\newblock In {\em Parallel and Distributed Systems, 2001. ICPADS 2001.
  Proceedings. Eighth International Conference on}, pages 165--170. IEEE, 2001.

\bibitem{bit1}
Mihai~F Ionescu and Klaus~E Schauser.
\newblock Optimizing parallel bitonic sort.
\newblock In {\em Parallel Processing Symposium, 1997. Proceedings., 11th
  International}, pages 303--309. IEEE, 1997.

\bibitem{bit2}
David Nassimi and Sartaj Sahni.
\newblock Bitonic sort on a mesh-connected parallel computer.
\newblock {\em IEEE Transactions on Computers}, 100(1):2--7, 1979.

\bibitem{s6}
Michael Hofmann and Gudula Runger.
\newblock A partitioning algorithm for parallel sorting on distributed memory
  systems.
\newblock In {\em High Performance Computing and Communications (HPCC), 2011
  IEEE 13th International Conference on}, pages 402--411. IEEE, 2011.

\bibitem{rad1}
Shin-Jae Lee, Minsoo Jeon, Dongseung Kim, and Andrew Sohn.
\newblock Partitioned parallel radix sort.
\newblock {\em Journal of Parallel and Distributed Computing}, 62(4):656--668,
  2002.

\bibitem{radix1}
Andrew Sohn and Yuetsu Kodama.
\newblock Load balanced parallel radix sort.
\newblock In {\em Proceedings of the 12th international conference on
  Supercomputing}, pages 305--312. ACM, 1998.

\bibitem{merge}
Martina-Cezara Albutiu, Alfons Kemper, and Thomas Neumann.
\newblock Massively parallel sort-merge joins in main memory multi-core
  database systems.
\newblock {\em Proceedings of the VLDB Endowment}, 5(10):1064--1075, 2012.

\bibitem{sample}
Hanmao Shi and Jonathan Schaeffer.
\newblock Parallel sorting by regular sampling.
\newblock {\em Journal of Parallel and Distributed Computing}, 14(4):361--372,
  1992.

\bibitem{sample2}
Peter Sanders and Sebastian Winkel.
\newblock Super scalar sample sort.
\newblock In {\em European Symposium on Algorithms}, pages 784--796. Springer,
  2004.

\bibitem{spark1}
Reynold Xin, Parviz Deyhim, Ali Ghodsi, Xiangrui Meng, and Matei Zaharia.
\newblock Graysort on apache spark by databricks.
\newblock {\em GraySort Competition}, 2014.

\bibitem{spark2}
Juwei Shi, Yunjie Qiu, Umar~Farooq Minhas, Limei Jiao, Chen Wang, Berthold
  Reinwald, and Fatma {\"O}zcan.
\newblock Clash of the titans: Mapreduce vs. spark for large scale data
  analytics.
\newblock {\em Proceedings of the VLDB Endowment}, 8(13):2110--2121, 2015.

\bibitem{Tim1}
Tim Peters.
\newblock Timsort, 2002.

\end{thebibliography}
\bibliographystyle{unsrt}

\end{document}